\documentclass[12pt]{article}
\usepackage{amsmath}
\usepackage{amssymb}
\usepackage{graphicx}
\usepackage{color}

\newcommand{\rf}[1]{(\ref{#1})}
\newcommand{\beq}{\begin{equation}}
\newcommand{\eeq}{\end{equation}}
\newcommand{\bea}{\begin{eqnarray}}
\newcommand{\eea}{\end{eqnarray}}

\newcommand{\e}{{\rm e}}

\newcommand{\plu}{\!+\!}

\newcommand{\lam}{\lambda}
\newcommand{\Lam}{\Lambda}

\newcommand{\Om}{\Omega}
\newcommand{\del}{\delta}

\newcommand{\sg}{\sigma}
\newcommand{\kp}{\kappa}

\newcommand{\dg}{\dagger}

\newcommand{\ra}{\rangle}

\newcommand{\prt}{\partial}

\newcommand{\cH}{{\cal H}}

\newcommand{\hH}{{\hat{H}}}

\begin{document}

\begin{center}
\vspace{24pt}
{ \Large \bf Late time acceleration of our Universe caused 
by the absorption of baby universes}\footnote{Contribution to the proceedings volume for {\it Gravity, Strings and Fields: A conference in honour of Gordon Semenoff.}}

\vspace{24pt}

{\sl J.\ Ambj\o rn}$\,^{a,b}$,
and {\sl Y.\ Watabiki}$\,^{c}$

\vspace{10pt}

{\small

$^a$~The Niels Bohr Institute, Copenhagen University\\
Blegdamsvej 17, DK-2100 Copenhagen \O , Denmark.\\
email: ambjorn@nbi.dk
\vspace{10pt}

$^b$~Institute for Mathematics, Astrophysics and Particle Physics
(IMAPP)\\ Radbaud University Nijmegen, Heyendaalseweg 135, 6525 AJ, \\
Nijmegen, The Netherlands

\vspace{10pt}

$^c$~Tokyo Institute of Technology,\\ 
Dept. of Physics, High Energy Theory Group,\\ 
2-12-1 Oh-okayama, Meguro-ku, Tokyo 152-8551, Japan\\
{email: watabiki@th.phys.titech.ac.jp}

}

\end{center}

\vspace{24pt}

\begin{center}
{\bf Abstract}
\end{center}

\noindent
If  our Universe is allowed to absorb baby universes,   
one obtains a  modified Friedmann equation that can 
explain the late time  acceleration of our Universe and there
 is no need for  a cosmological constant. In addition the modified Friedmann
 equation favors the value of the Hubble constant obtained by local 
 measurements.

\newpage

\section{Introduction}

Causal Dynamical Triangulations (CDT) is an attempt to define 
a non-perturbative theory of quantum gravity (see \cite{physrep,lollreview} 
for reviews). It is formulated in a proper-time gauge and one can explicitly 
perform a rotation to Euclidean signature where the integrand in the
path integral is then changed from $e^{iS_L}$ to $e^{-S_E}$, the 
Euclidean action $S_E$ being real. This implies
that the four-dimensional CDT theory can be addressed by Monte Carlo
simulations. One of the  outcomes of these simulations is 
that when the topology of space is $T^3$ 
one observes the following effective action as a function 
of the three-volume $V(t)$ at proper-time $t$:
\beq\label{j1}
 S=  \frac{1}{\Gamma}\int dt  \, \Big( \frac{\dot{V}^2}{2 V} + \Lambda  V\Big).
 \eeq
 This is remarkable, because it is essentially the Hartle-Hawking 
 minisuperspace action (including the  rotation of the conformal factor) \cite{hh}.
 The rotation of the conformal factor was proposed as a solution 
 to the  problem of the unboundedness from below of the Euclidean 
 Einstein-Hilbert action. Also CDT solves the problem, but in CDT the 
 effective action arises by integrating out (via the Monte Carlo 
 simulations) all other degrees of freedom than $V(t)$, while Hartle and Hawking by hand restricted the geometry to only depend on $V(t)$.
 
 Recall the standard minisuperspace approximation, where one uses the 
 metric
  \beq\label{j3}
 ds^2 = -N^2(t)dt^2 + a^2(t) d\Om_3,\qquad d \Om_3 = \sum_{i=1}^3 dx_i^2.
 \eeq
 Introduce 
 \beq\label{j3a} 
 v(t) = \frac{1}{\kp} a^3(t),\quad \kp = 8\pi G,  
 \eeq
 where $G$ the gravitational constant.
Then the minisuperspace Einstein-Hilbert action is 
 \beq\label{j2}
 S =  \int dt  \,\Big(-\frac{\dot{v}^2}{3Nv} - \lam N v\Big),
 \eeq
 where $\lam$ is the cosmological constant. 
 The  Hubble parameter $H(t)$ is defined as 
 \beq\label{j3b}
 H(t) \equiv \frac{\dot{a}}{a} = \frac{1}{3} \, \frac{\dot{v}}{v}.
 \eeq
 The Hamiltonian corresponding to \rf{j2} is 
\beq\label{j4}
\cH(v,p)  = N v \Big(- \frac{3}{4} p^2+ \lam\Big),
\eeq
where $p$ denotes the momentum conjugate to $v$. 
Variation with respect to $N(t)$ ensures that the 
classical solutions are ``on shell'', i.e. $\cH(v,p)=0$ for a solution.
In the following we will for simplicity gauge fix $N(t) =1$, but 
only consider classical solutions corresponding to  $\cH(v,p)$ that 
satisfy $\cH(v,p)=0$. 
Such a  classical solution is  de Sitter spacetime where  
\beq\label{j5}
 v(t) = v(t_0) \, \e^{ \sqrt{ 3 \lam} \,t}.
 \eeq
 The analytic continuation of \rf{j2} by Hartle and Hawking, including the 
 rotation of the conformal factor, will result in the following action:
  \beq\label{jx2}
 S_{hh} =  \int dt  \,\Big(\frac{\dot{v}^2}{3Nv} + \lam N v\Big),
 \eeq
Hartle and Hawking were mainly interested in using \rf{jx2}
in the path integral. Whatever result one would obtain, one would 
eventually have to rotate back to Lorentzian signature, if one is 
interested in cosmological applications, and if the quantum theory 
has a classical limit, this limit  should be given by \rf{j2} and \rf{j4}. Thus 
we will not expect the late time aspects of cosmology to be directly affected by the 
quantum aspects of gravity. Similarly, since CDT is intended to be 
a quantum gravity theory, the result \rf{j1} indicates CDT will not provide 
us with new insight about the late time cosmology. ``Traditional'' 
quantum gravity might affect the early time universe (cure Big Bang singularities
etc.), not the late time universe. 
However, if we allow for some more ``untraditional'' quantum phenomena  
like the absorption and emission of so-called baby-universes, 
this situation can change\footnote{Similar ideas have recently been advocated in \cite{newkawai}.}

\section{A Universe that absorb and emit baby universes}
  
 By allowing our Universe to absorb and emit baby universes, we are 
 really dealing with a multi-verse theory. In order to be able to 
 perform some calculation in such a multi-verse theory, we will 
 work in a minisuperspace approximation where the spatial universe
at a give proper time $t$ is characterized by its spatial volume $v(t)$. 
In a corresponding quantum theory we will denote a state with 
spatial volume $v$ by $ | v \ra$. Let now $\hat{\cH}$ denote 
the quantum mechanical Hamiltonian corresponding 
to the action $S_{hh}$ given by \rf{jx2}:
\beq\label{j13}
\hat{\cH}^{(0)}=  
{v}\Big(-  \frac{3}{4} \frac{d^2}{d v^2}\,+\lam  \Big).
\eeq
This can now be viewed as a single universe Hamiltonian and 
 it was already encountered in the study of 
two-dimensional CDT \cite{al,al1}. There is a well defined Hilbert space 
corresponding to $\hat{\cH}^{(0)}$ and the multi-verse Hilbert space will
now be the Fock space constructed from the single 
universe states $| v \ra$. A natural many-universe Hamiltonian that 
allows for a universe to split in two or two universes to merge to a single universe is then 
\bea
\hH &=& \hH^{(0)} - g \int dv_1 \int dv_2 \;\Psi^\dg(v_1)\Psi^\dg(v_2)\;
(v_1\plu v_2)\Psi(v_1\plu v_2) 
\label{j12} \\ 
&& \!\!\! -g\int dv_1 \int dv_2 \;\Psi^\dg(v_1\plu v_2)\;v_2\Psi(v_2)\;v_1\Psi(v_1)
- \!\int \frac{dv}{v}\,  \rho(v) \Psi^\dg (v), \nonumber
\eea
where 
\beq\label{j13}
\hH^{(0)} = \int_0^\infty \frac{dv}{v} \; \Psi^\dg (v) 
\hat{\cH}^{(0)} \, v  \Psi(v),\qquad \rho(v)= \del(v),
\eeq
and where  $\Psi^\dg(v)$ and $\Psi(v)$ are creation and annihilation operators
 for single universes of spatial volume $v$. 
 The existence of the last term in eq.\ \rf{j12} implies that a universe can be 
created from the Fock vacuum $| 0 \ra$ if the spatial volume is zero and 
therefore the formal Fock vacuum is not stable.

\begin{figure}[t]
\centerline{\scalebox{0.2}{\rotatebox{0}{\includegraphics{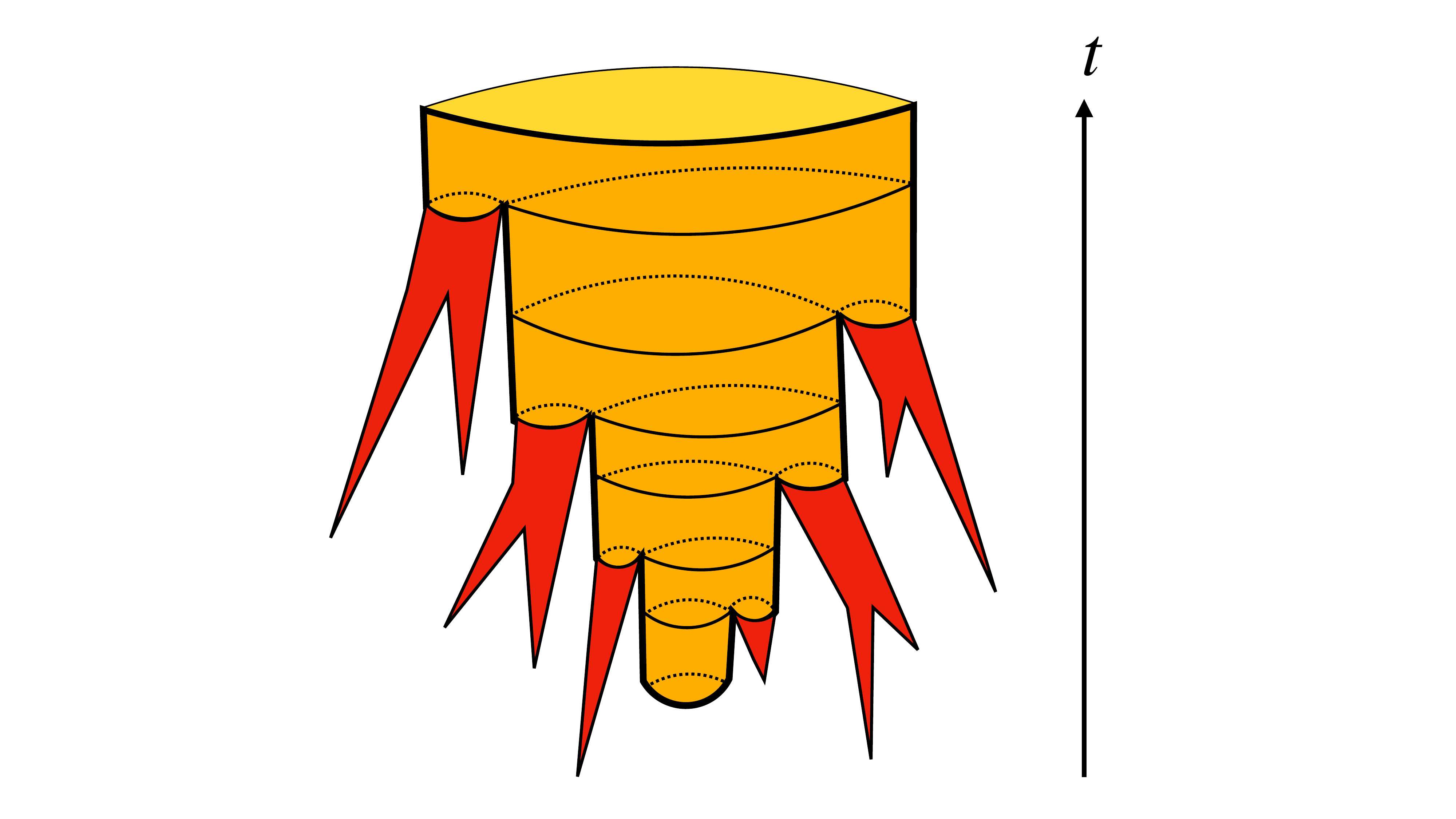}}}}
\caption[figxj1]{{\small
Our Universe (orange), represented as  one-dimensional circles, 
propagating in time, with baby universes (red) merging and increasing the spatial volume.
}}
\label{figxj1}
\end{figure}

 The Hamiltonian \rf{j12} was introduced  in \cite{sft} and there exists 
 a truncation, denoted generalized 2d CDT (GCDT) that can be solved 
 analytically \cite{gcdt,matrixgcdt}. 
  It follows the evolution of our  Universe  in time and allows 
  other universes (called baby universes) 
  to merge with our universe during this time 
  evolution, but does not allow our Universe to split in two. 
  This is illustrated in Fig.\ \ref{figxj1}.
  The effective Hamiltonian   (the so-called 
  inclusive Hamiltonian first introduced in  \cite{inclusive})  can 
 be found from \rf{j12} as follows: replace   the quantum field $\Psi(v)$, representing the disappearance of a universe of spatial volume $v$ by $\psi(v) +\Psi(v)$, where $\psi(v)$ 
 is viewed as a classical field representing the distribution of the spatial 
 volumes of the baby universes being absorbed. Let 
 $F(p)$ denote  the Laplace transform of $\phi(v) = v \psi (v)$:
\beq\label{j35}
 \phi(v) = \phi_0 + \phi_1{v}+ \cdots, \quad
F(p) =\!\! \int_0^\infty \!\!\!dv \;\e^{-p \,v} \phi(v) = \frac{\Gamma(1)\phi_0}{ p} + 
\frac{\Gamma(2)\phi_1}{ p^2} + \cdots 
\eeq
After some algebra one obtains for the quadratic part of $\hH$ that determines 
the propagation of the universe:
\beq\label{j35a}
\hH_{\rm eff} = \hH^{(0)} - 
2g\int dv \Psi^\dg(v)\, F\!\left(\frac{d}{dv}\right) v \Psi(v),
\eeq
and we can write
\beq\label{j36}
\hat{\cH}_{\rm \rm eff} = \hat{\cH}_0 -2g  F\!\left( \frac{d}{dv}\right) {v}.
\eeq

In GCDT $F(p)$ is determined by the self-consistency   requirement that our Universe, modified by the impact $\phi(v)$ of  baby universes,  should be identical to the 
baby universes it absorbs.  We refer to \cite{gcdt} (or the book 
\cite{book2}) 
for the details of this determination. We will allow for more general $F(p)$.
To obtain the classical effective Hamiltonian relevant for cosmology
we have to make the analytic rotation  back from the 
Hartle-Hawking metric  and replace  $ -id/dv$ by the 
classical momentum $p$ conjugate to $v$. In this way we obtain
\beq
\cH_{\rm eff} =  {v}\Big( - \,\frac{3}{4} p^2 + {\lam} -2 g F( p)\Big) = 
-v f(v).
\label{j38}
\eeq
The on shell solution $v(t)$ satisfies
\beq\label{jx2a}
\dot{v} = \frac{\prt \cH_{\rm eff}}{\prt p} = -v f'( p),\quad f(p) = 0.
\eeq
This implies that $p$ is constant and one has (for suitable $f(v)$) 
an exponentially growing solution, just like the de Sitter solution 
\rf{j5}. In particular we can have an exponentially growing solution 
even if the cosmological constant $\lam =0$. In such a case the 
exponential growth is caused by the impact of other universes on 
our Universe, as illustrated in Fig.\ \ref{figxj1}. 

In the following we will consider the simplest choice of $F(p)$, 
namely the one corresponding the $\phi (v)= \phi_0$. This is 
the limit of any $\phi(v)$ where the impact volume of the 
baby universes is zero, so with this choice it is really appropriate 
to denote the incoming universes ``baby'' universes. In addition we 
will assume that the cosmological constant is zero. Choosing for 
convenience $\phi_0 = 3/4$ we can now write
\beq\label{jx3}
\cH_{\rm eff}  = - v f(v), \quad f(p) = \frac{3}{4} \Big( p^3 + \frac{2g}{p} \Big),
\eeq
and we find the  exponential growth
\beq\label{jx4}
v(t) = v(t_0) \, \e^{\frac{3}{2} (2g)^{1/3}  t}.
\eeq

 \section{Including CDM matter}
 
 We now include matter in our cosmological model.
 Since we are only interested in the late time cosmology,
 we include it as a CDM density $\rho_{\rm m}(v)$ 
 added to the Hamiltonian \rf{jx3}. We do not include the matter 
 density in the baby universes, since they are precisely baby universes 
 with infinitesimal volume. 
 Thus the 
Hamiltonian has the form
\beq\label{j30}
 \cH [v,p] =  {v} \,( -f( p)+ \kp \rho_{\rm m}(v) \,), \quad  
 f(p) = \frac{3}{4} \Big( p^2 + \frac{2g}{p} \Big),
 \quad v \rho_{\rm m}(v) = v_0 \rho_{\rm m} (v_0), 
\eeq
where $v_0$ and $\rho_{\rm m}(v_0)$ denote the values at the present time $t_0$.  Let us 
discuss the solution of the eoms for arbitrary $f( p)$ in \rf{j30}. The 
eoms simplify since  $v \rho_{\rm m}(v)$ is constant.  
\beq\label{j41}
\dot{v} = \frac{\prt \cH}{\prt p} = -v f'( p),\quad {\rm i.e.} \quad  3 \,\frac{\dot{a}}{a} =
\frac{\dot{v}}{v} = -f'( p),
\eeq
\beq\label{j42}
~~\dot{ p} = - \frac{\prt \cH}{\prt v } =  f( p), \quad {\rm i.e.} \quad 
t = \int_{ -\infty}^{ p} \frac{d \,  p}{f( p)} \, .
\eeq
By construction any solution to \rf{j41}-\rf{j42} will satisfy $\cH = {\rm const}$, and we are 
interested in the ``on-shell'' solutions $\cH=0$, which by \rf{j30} implies that 
\beq\label{j40}
f (p) = \kp \rho_{\rm m}(v) = \kp \rho_{\rm m}(v_0) \frac{ v_0}{v} =
f( p_0) \frac{ v_0}{v} = f( p_0)(1+z)^3, 
\eeq
where $p_0$ denotes the value of $p$ at present time $t_0$ and 
$z$ denotes the redshift at time $t$, i.e. 
\beq\label{j40a}
z(t)+1 = \frac{a(t_0)}{a(t)} = \Big(\frac{v(t_0)}{v(t)}\Big)^{1/3}.
\eeq
{\it Eq.\ \rf{j40} is the generalized Friedmann when eq.\ \rf{j41} is used to express 
$ p$ in terms of ${\dot{a}}/{a}$.}  In \cite{aw2,aw2a,aw3} 
we studied this Friedmann 
equation for $f(p)$ given by \rf{j30} and called it {\it the modified Friedmann 
equation}. 
We will restrict ourselves to that case in the following,  and  
one can then integrate \rf{j42} analytically to find 
\beq\label{ap12}
t =\int_{-\infty}^{ p} \frac{d  p'}{f( p')} =
\frac{4}{3 \sqrt{3}\, a}\Big( \arctan \frac{2 p -a}{\sqrt{3}\, a} +  
\frac{1}{2\sqrt{3} }\log \frac{ ( p-a)^2 +a p}{(p+a)^2}   + \frac{ \pi}{2}\Big),
\eeq
where $a= (2g)^{1/3} $. A more general 
study for other functions $f(v)$, including the  $f(p)$ corresponding to GCDT, 
can be found in \cite{aw4}.

\section{Comparison with late cosmological data}

Our model has two  coupling constants 
coupling constant,  $\kp$ and $g$. We consider $\kp$ fixed by local experiments and we will not discuss it any further. If our model 
should describe late cosmology well, the coupling constant $g$ 
 has to be quite small. The situation
is the same as for the cosmological constant in the standard $\Lam$CDM
model, where the cosmological constant $\lam$ also must be small.
In both cases they will play no role in the time development  
of the Universe for  $t < t_{\rm LS}$, the time of last scattering. 
Thus we can calculate the time of last scatting $t_{\rm LS}$
using standard cosmology without any reference to $g$ (or $\lam$).
Knowing $t_{\rm LS}$ we can obtain $p_{\rm LS}$ also 
without any reference to $g$ (or $\lam$) from
\beq\label{j63}
t_{\rm LS} =  \int_{-\infty}^{ p_{\rm LS}}   \frac{d  p}{\frac{3}{4}\, p^2 + 
\frac{1}{3} \,\kp \rho_{\rm r}(v( p))}, \qquad \frac{3}{4}\, p^2 = 
\kp \rho_{\rm m}(v) + \kp \rho_{\rm r}(v).
\eeq
We have included in the formula the radiation density, since that 
cannot be ignored all the way down to $t=0$ in the 
region $t <  t_{\rm LS}$. This formulas allows 
us to write  $p_{\rm LS}(t_{\rm LS})$,
still without any reference to $g$, and finally we can also write 
\beq\label{jx10}
f(p_{\rm LS}) = \frac{3}{4} p_{\rm LS}^2,
\eeq
since $p_{\rm LS}^2 \gg g/|p_{\rm LS}|$ for $t \leq t_{\rm LS}$.
For $t >  t_{\rm LS}$  we  expect that the 
CDM  $\rho_{\rm m} (v)$ will be a good approximation to the matter density,
and under this assumption  $g$ and $t_0$ are    
determined  from the values the  $H_0$ and $z_{\rm LS}$, the red shift 
at the time of last scattering. This is 
interesting since $H_0$ and $z_{\rm LS}$ 
can be obtained by measurements that are almost independent 
of cosmological models\footnote{More precisely $H_0$ can be (and is) determined by 
local measurements. This is the value we denote $H_0^{\rm SC}$ below. Similarly,
 observations  allow us to determine  the temperature 
$T(t_0)$ of the CMB. $T(t_{\rm LS})$ can be calculated by atomic physics and is to a large 
extent independent of the cosmological model, as is also the statement that 
$T(t_{\rm LS})/T(t_0) = a(t_0)/a(t_{\rm LS}) = 1+z_{\rm LS}$.}.

Given  $H_0$ and $z_{\rm LS}$ we can first obtain  $p_0$ ( $p$ at $t_0$, our present time)
 for a given value of $g$ from \rf{j41}
\beq\label{j44}
f'( p_0) = - 3 H_0\, .
\eeq
Next the Friedmann equation \rf{j40} applied at $p= p_{\rm LS}$ reads
\beq\label{j44a}
 f( p_0)=  \frac{f( p_{\rm LS})}{(1+z_{\rm LS})^3} = 
 \frac{ 3 p_{\rm LS}^2}{4 (1+z_{\rm LS})^3}\, .
 \eeq
 Since we know $p_{\rm LS}$ from \rf{j63} and we know $f(p_0)$ as a function
 of $g$, this Friedmann equation will determine $g$.
Finally $t_0-t_{\rm LS}$ is the determined from \rf{j42}  (or  \rf{ap12} for the
the specific case of the modified Friedmann equation).

We have detemined the coupling constant $g$ and $t_0$ 
from $H_0$  and $z_{\rm LS}$. There are 
presently  two values of $H_0$ that do not agree within $5\sg$ (the 
so-called $H_0$ tension). One value
is deduced from ``local'' measurements, using various ``space candles'' (SC) 
techniques \cite{h0sc}. We denote 
it $H_0^{\rm SC}\,(= 73.04 \pm 1.04 \;{\rm km/s/Mpc})$, and it 
is almost independent of cosmological models. 
The other value is deduced from the CMB data created at 
 $t_{\rm LS}$. It is using  cosmological models in a number of ways, among 
those  to extrapolate to present time $t_0$. This value  we denote
 $H_0^{\rm CMB}$ and it is model dependent. The value $H_0^{\rm CMB}= 67.4 \pm 0.5\;{\rm km/s/Mpc}$  quoted in \cite{planck}, and 
the one  we will use here, refers the $\Lam$CDM model.

\begin{figure}[t]
\centerline{
\scalebox{0.8}{\rotatebox{0}{\includegraphics{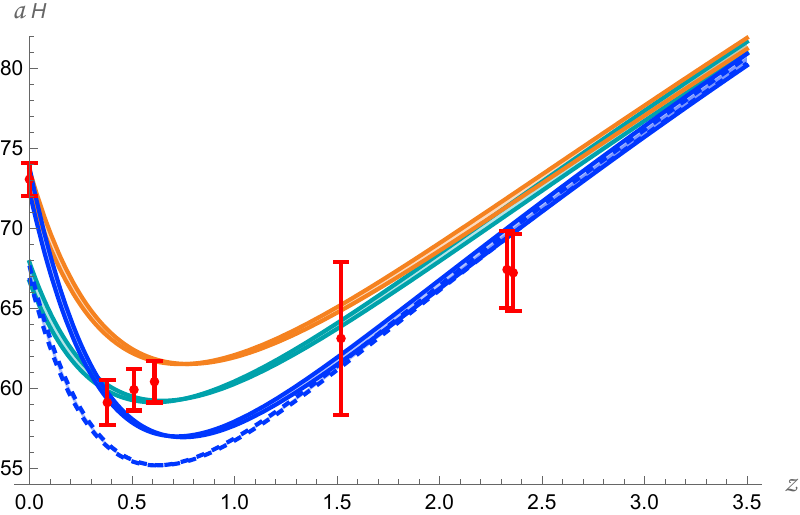}}}}
\caption[fig3]{{\small
The blue curves show $H(z) a(z)$, $a(z) =1/(1+z)$, 
for the modified Friedmann equation starting from  $H_0^{\rm SC}$ and 
$H_0^{\rm CMB}$ (the dotted blue curve). The  
green  curve  shows $H(z) a(z)$  based on the 
$\Lambda$CDM model with $H_0^{\rm CMB}$  and $\lam^{\rm CMB}$ (the cosmological constant) taken
from \cite{planck}. The orange curve shows $H(z)a(z)$  based on the 
$\Lambda$CDM model with  $\lam$ chosen such that it starts with $H_0^{\rm SC}$.
Inserted in red are (apart from the SC data at $z \approx 0$)
also independent data from other observations: first three points from the baryon acoustic oscillation data \cite{BAO}, 
the next from quasars \cite{quasars} and the last two data points from Ly--$\alpha$ measuments \cite{lyalpha1,lyalpha2}. }}
\label{fig3}
\end{figure}

A good check of our cosmological model is  to compare its low $z$ (late time) predictions
with (almost model independent) low $z$ measurements. This is done in Fig.\ \ref{fig3}. 
 It is seen that our modified Friedmann equation agrees very well 
 with  the small $z$ data if we use $H_0^{\rm SC}$ and not well
 if we use $H_0^{\rm CMB}$. By contrast the $\Lam$CDM model 
 agrees well with the data when we use $H_0^{\rm CMB}$ and not well 
 when we use $H_0^{\rm SC}$. In Table 1 we have listed the reduced 
 $\chi^2$ values, $\chi^2_{\rm red}(\Lam )$ for the $\Lam$CDM model and
 $\chi^2_{\rm red}({\rm MF})$ for the modified Friedmann (MF) equation,
obtained by comparing the calculated $H(z)$ with the observe $H(z)$. 
Finally, in Table 2 we  list the calculated values of the present days time $t_0$  in the four cases.
 
 \vspace{6pt}
  
 \begin{center}
 \begin{tabular}{|c|c|c|} \hline
 ~& $\chi^{2}_{\rm red}(\Lam )$ & $\chi^2_{\rm red}$(MF)\\
 \hline
  $H_0^{\rm SC}$ & 3.5 & 1.8\\
 \hline
   $H_0^{\rm CMB}$ & 1.2  & 5.6\\
 \hline
 \end{tabular}\\
 
 \vspace{6pt}
 Table 1.
 \end{center}

\vspace{3pt}

\begin{center}
\begin{tabular}{|c|c|c|} \hline
 ~& $t_0(\Lam )$ & $t_0$(MF)\\
 \hline
  $H_0^{\rm SC}$ & 13.3~{\rm  Gyr}   & 13.9~ {\rm Gyr}\\
 \hline
   $H_0^{\rm CMB}$ & 13.8~ {\rm  Gyr} & 14.4~ {\rm Gyr} \\
 \hline
 \end{tabular}\\

 \vspace{6pt}
 
 Table 2.
 \end{center}

 We conclude that if $H_0^{\rm CMB}$ should be proven to be the 
 correct value, our modified Friedmann equation is ruled out. Similarly,
 should $H_0^{\rm SC}$ turn out to be correct, the $\Lam$CDM model
 seems to have a problem, while the modified Friedmann equation seems
 to do well. 
 
 The density fluctuations of matter is another  ``local'' observable
 that will be available for  $z < 2$.
 Presently the error bars on these data  are too large to provide 
 a serious testing of the modified Friedmann equation. 
 However, this will most likely 
change dramatically with the new observations to be obtained by  the Euclid satellite.

 \section{Discussion}
 
 We have proposed a modified Friedmann equation that results 
 in a late time cosmology that is different from the late time cosmology 
 provided by the $\Lam$CDM model. It does not require 
 a cosmological constant, the late time acceleration  of 
 our Universe being 
 generated by the bombardment of  the Universe 
 by baby universes. The model fits the low $z$ data very well provided 
 that the $H_0^{\rm SC} = 73.04 \pm 1.04$ km/s/Mpc is the correct value of 
 the Hubble constant.
 
 The model has a formal equation of state parameter  $w(z) < -1$. 
 In standard cosmology this is a sign that
some unphysical degrees of freedom have been added to the system.
However, in the modified model this is not the case. In fact the 
term $(-3g/2p)\cdot v$ added to the effective Hamiltonian is more like a 
time dependent cosmological constant term, but without usual 
problem that a time dependent cosmological constant will break the 
invariance under time-reparametrization. An increasing time dependent cosmological will result in $w(z) < -1$ in an expanding universe. 
In our model   $-3g/2p$ is an increasing function of $t$ since $p(t)$ 
increases from large negative values towards the value $-(2g)^{1/3}$
for $t \to \infty$ where it will act as an ordinary cosmological term 
with value $3 (2g)^{2/3}/4$. $w(z)$ changes monotonically 
from $-3/2$ at $z=\infty$ ($t =0$) to $-1$ for $z=-1$ ($t= \infty$).

Before  observations showed that the cosmological constant was positive 
(but very small), many favored that $\lam =0$, maybe caused 
by some not fully understood dynamics, like  the one offered by 
Coleman's mechanism \cite{coleman}. In some sense it might be easier 
to explain why $\lam =0$ than to explain the very small value observed today.
What we have argued here is that even if one can show that $\lam =0$, we
can still have an accelerating late universe, the acceleration due to baby
universe absorption. In order to fit observations our coupling constant $g$ also
has to be very small. However, it appears in a larger multi-verse theory and 
one could hope that being able to solve this theory might also help us 
understanding the smallness of $g$. The program to understand this 
theory has been started  \cite{aw1,aw1a,aw1b}, but it is still work in progress.

%
%

\end{document}